\documentstyle[12pt]{article}
\begin{document}    
\setcounter{page}{1}
\thispagestyle{empty}
\begin{flushright}    
hep-ph/0101334\\
January, 2001
\end{flushright}   
\begin{center}
\vskip 10mm
{\large\bf Possible Flavor Mixing Structures of Lepton Mass Matrices}
\vskip 15mm
{\bf Naoyuki Haba\footnote{E-mail address:
    haba@eken.phys.nagoya-u.ac.jp}}
\vskip 1mm
{\it Faculty of Engineering, Mie University, Tsu, Mie 514-8507,
  Japan}
\vskip 3mm
{\bf Joe Sato\footnote{E-mail address: joe@rc.kyushu-u.ac.jp}}
\vskip 1mm
{\it Research Center for Higher Education, Kyushu University, Fukuoka
  810-8560, Japan}
\vskip 3mm
{\bf Morimitsu Tanimoto\footnote{E-mail address:
    tanimoto@muse.hep.sc.niigata-u.ac.jp}}
\vskip 1mm
{\it Department of Physics, Niigata University, Niigata 950-2181,
  Japan}
\vskip 2mm
and
\vskip 2mm
{\bf Koichi Yoshioka\footnote{E-mail address:
    yoshioka@yukawa.kyoto-u.ac.jp}}
\vskip 1mm
{\it Yukawa Institute for Theoretical Physics, Kyoto University, Kyoto
  606-8502, Japan}

\vskip 18mm
{\bf Abstract}
\end{center}
To search for possible textures of lepton mass matrices, we
systematically examine flavor mixing structures which can lead to
large lepton mixing angles. We find out 37 mixing patterns are
consistent with experimental data, taking into account phase factors
in the mixing matrices. Only six of the patterns can explain the
observed data without any tuning of parameters, while the others need
particular choices for the phase values. It is found that these six
mixing patterns are those predicted by the models which have been
proposed to account for fermion mass hierarchies. On the other hand,
the others may give new flavor mixing structures of lepton mass
matrices and therefore new possibilities of model construction.

\newpage
\baselineskip=22pt
\topskip 0mm
\renewcommand{\thefootnote}{\fnsymbol{footnote}}
\setcounter{footnote}{0}

\section{Introduction}

The Super-Kamiokande experiment has confirmed neutrino oscillations 
in the atmospheric neutrinos, which favors 
the $\nu_\mu\rightarrow\nu_\tau$ process with a large mixing 
angle $\sin^2 2\theta_{\rm atm} \geq 0.88$ and mass-squared difference
$\Delta m^2_{\rm atm}=(1.6-4)\times 10^{-3}$ eV$^2$~\cite{SKam}. On
the other hand, for the solar neutrino problem~\cite{SKamsolar}, the
recent data of Super-Kamiokande seems to favor the large mixing angle
(LMA) Mikheyev-Smirnov-Wolfenstein (MSW) solution~\cite{N2000}, but
four solutions are still experimentally allowed; small mixing angle
(SMA) MSW~\cite{MSW}, LMA-MSW, low $\Delta m^2$ (LOW), and vacuum
oscillation (VO) solutions~\cite{BKS}. As a result, the neutrino
mixing matrix (Maki-Nakagawa-Sakata (MNS) matrix~\cite{MNS}) has two
possibilities: one is the matrix with single maximal mixing, which
gives the SMA-MSW solution for the solar neutrino problem, and the
other with bi-maximal mixing~\cite{bimax}, which corresponds to the
LMA-MSW, LOW, and VO solutions.

Assuming that the neutrino oscillations account for the solar and
atmospheric neutrino data, one can consider the prototypes of the MNS
mixing matrix $U_{\rm MNS}$ which are written as
\begin{eqnarray}
  U_{\rm MNS} \,\simeq\, \left( 
    \matrix{ 1 & 0 & 0 \cr
      0 &  \frac{1}{\sqrt{2}} & \frac{1}{\sqrt{2}} \cr
      0 & -\frac{1}{\sqrt{2}} & \frac{1}{\sqrt{2}} \cr } \right),
  \qquad\quad
  \frac{\Delta m_\odot^2}{\Delta m^2_{\rm atm}}\,\simeq\, 10^{-2},
\label{single}
\end{eqnarray}
for single maximal mixing, and 
\begin{eqnarray}
  U_{\rm MNS} \,\simeq\, \left( 
    \matrix{ \frac{1}{\sqrt{2}} & \frac{1}{\sqrt{2}} & 0 \cr
      -\frac{1}{2} &  \frac{1}{2} & \frac{1}{\sqrt{2}} \cr
      \frac{1}{2} & -\frac{1}{2} & \frac{1}{\sqrt{2}} \cr } \right), 
  \qquad \frac{\Delta m_\odot^2}{\Delta m^2_{\rm atm}}\,\simeq\,
  \left\{
    \matrix{10^{-2} \ \ (\mbox{LMA--MSW}), \hfill\cr 
      10^{-4} \ \ ({\rm LOW}), \hfill\cr
      10^{-6} \ \ ({\rm VO}), \hfill\cr } \right.
\label{bi}
\end{eqnarray}
for bi-maximal mixing. Here $\Delta m_\odot^2$ is the mass-squared
difference relevant to the solar neutrino problem.

To clarify the origins of these nearly maximal mixings is one of the
most important issues in flavor physics. In constructing the models for
fermion masses and mixing, there are some preferred bases given by
underlying theories, such as grand unified theories (GUT). For the MNS
matrix in Eq.~(\ref{single}), the maximal mixing angle may follow from
the charged-lepton mass matrix, the neutrino mass matrix, or both of
them, depending on the models under consideration. In the case of
bi-maximal mixing in Eq.~(\ref{bi}), the situation is more
non-trivial. It is therefore important in light of model construction
to search for possible mixing patterns of charged leptons and
neutrinos. In this paper, we systematically investigate the mixing
patterns where at least one of the mixing matrices has sources of
maximal mixing. As we will see, our analyses are independent of
particular structures of lepton mass matrices and hence of the mass
spectrum of neutrinos. The results are also not concerned with whether
the neutrinos are Majorana or Dirac particles, in other words, whether
the right-handed neutrinos exist or not. Based on our results, we
discuss new possibilities of the forms of lepton mass matrices, which
may account for the experimental data.

In section 2, we discuss the mixing patterns of charged leptons and
neutrinos, and classify them in light of the phenomenological
constraints from Super-Kamiokande and long baseline neutrino
experiments. In section 3, we show several examples of mass matrices
of charged leptons and neutrinos that can give the allowed mixing
patterns obtained in section 2. Section 4 is devoted to summary and
discussions.

%%%%%%%%%%%%%%%%%%%%%%%%%%%%%%%%%%%%%%%%%%%%%%%%%%%%%%%%%%%%%%%%%%%%%
%%%%%%%%%%%%%%%%%%%%%%%%%%%%%%%%%%%%%%%%%%%%%%%%%%%%%%%%%%%%%%%%%%%%%
\section{Phenomenology of Mixing Matrices}

In this section, we study possible flavor mixing structures of
leptons, which can lead to large mixing angles. Given the
charged-lepton and neutrino mass matrices, the MNS mixing matrix is
defined as
\begin{equation}
  U_{\rm MNS} \,=\, V_E^\dagger\, V_\nu,
\end{equation}
where the $3\times 3$ matrix $V$'s are the mixing matrices which
rotate the left-handed fields so that the mass matrices are
diagonalized. The matrices $V_E$ and $V_\nu$ are generally
parametrized as follows:
\begin{eqnarray}
  V_E=P\,U(23)\,P'\,U(13)\,U(12)P'', \qquad
  V_\nu=\overline P\,\overline U(23)\,\overline P'\,\overline U(13)\,
  \overline U(12)\,\overline P''.
\end{eqnarray} 
Here $U(ij)$ are the rotation matrices,
\begin{equation}
  U(23)= \left( 
    \matrix{ 1 & 0 & 0 \cr
      0 & c_{23} & s_{23} \cr 
      0 & -s_{23} & c_{23} \cr} \right), \ \
  U(13)= \left( 
    \matrix{ c_{13} & 0 & s_{13} \cr
      0 & 1 & 0 \cr
      -s_{13} & 0 & c_{13} \cr} \right), \ \
  U(12)= \left( 
    \matrix{ c_{12} & s_{12} & 0 \cr
      -s_{12} & c_{12} & 0 \cr
      0 & 0 & 1 \cr} \right),
\end{equation}
where $s_{ij}=\sin\theta_{ij}$ and $c_{ij}=\cos\theta_{ij}$, 
and the $P$'s are the phase 
matrices; $P={\rm diag} (1, e^{ia}, e^{ib})$, 
$P'={\rm diag} (1, 1, e^{i\delta})$, and 
$P''={\rm diag} (e^{ip}, e^{iq}, e^{ir})$. The matrices with bars,
$\overline U(ij)$, $\overline P$, $\overline P'$, and $\overline P''$
in the neutrino side take the same parametrizations as above. In fact,
the phase factors in $P''$ are physically irrelevant in that they can
be absorbed with a redefinition of charged-lepton fields. For the
phases in $\overline P''$, the same prescription can be done in the
case of Dirac neutrinos, while for Majorana neutrinos these phases
cannot be absorbed into the neutrino fields, and remain physical. Note
however that they are irrelevant to the values of mixing angles and
hence we can safely drop the phase matrices $P''$ and $\overline P''$
in the following analyses. Now, the MNS matrix is given by
\begin{equation}
  U_{\rm MNS} \,=\, U_E^\dagger\,Q\,U_\nu,
\end{equation} 
where
\begin{eqnarray}
  U_E=U(23)\,P'\,U(13)\,U(12), &&
  U_\nu=\overline U(23)\,\overline P'\,\overline U(13)\,
  \overline U(12), \nonumber\\[3mm]
  Q = P^*\overline P &\equiv& \left( 
    \matrix{ 1 & 0 & 0 \cr
      0 & e^{i\alpha} & 0 \cr
      0 & 0 & e^{i\beta} \cr} \right).
  \label{Q}
\end{eqnarray} 
In $U_{\rm MNS}$, there are four phase parameters to be considered:
$\alpha$, $\beta$, $\delta_E$, and $\delta_\nu$. As will be seen below, 
in our analysis, the phase factors in the matrix $Q$ sometimes play
important roles to have phenomenologically viable mixing angles. The
mixing matrices $U(ij)$ and $\overline U(ij)$ are fixed when the mass
matrices of charged leptons and neutrinos are given in a concrete
model. On the other hand, from the view of mixing angles, there are
six mixing parameters in $U_E$ and $U_\nu$, and it is meaningful to
raise a query about which angles are responsible for the observed
maximal mixings in $U_{\rm MNS}$. In order to study this, we
phenomenologically analyze the mixing structures of lepton flavor
without referring to specific models.

In the first approximation, we assume that mixing angles are zero or
large, and examine possible combinations of $U_E$ and $U_\nu$
referring to the indications of Super-Kamiokande and long baseline
neutrino experiments. Let us consider the following nine types of
mixing matrices for $U_E$ and $U_\nu$. The first three types of
matrices are given by taking one of the mixing angles as maximal and
the others as zero:
\begin{eqnarray}
  A &=& \left( 
    \matrix{ 1 & 0 & 0 \cr
      0 & \frac{1}{\sqrt{2}} & \frac{1}{\sqrt{2}} \cr
      0 & -\frac{1}{\sqrt{2}} & \frac{1}{\sqrt{2}} \cr} \right), 
  \quad\qquad 
  \matrix{ s_{12}=0 \hfill\cr 
    s_{13}=0 \hfill\cr
    s_{23}=1/\sqrt{2} \hfill\cr} \\[3mm]
  S &=& \left( 
    \matrix{ \frac{1}{\sqrt{2}} & \frac{1}{\sqrt{2}} & 0 \cr
      -\frac{1}{\sqrt{2}} & \frac{1}{\sqrt{2}} & 0 \cr
      0 & 0 & 1 \cr} \right), 
  \quad\qquad 
  \matrix{ s_{12}=1/\sqrt{2} \hfill\cr 
    s_{13}=0 \hfill\cr
    s_{23}=0 \hfill\cr} \\[3mm]
  L &=& \left(
    \matrix{ \frac{1}{\sqrt{2}} & 0 & \frac{1}{\sqrt{2}} \cr
      0 & 1 & 0 \cr
      -\frac{1}{\sqrt{2}} & 0 & \frac{1}{\sqrt{2}} \cr} \right), 
  \quad\qquad
  \matrix{ s_{12}=0 \hfill\cr
    s_{13}=1\sqrt{2} \hfill\cr
    s_{23}=0 \hfill\cr}
\end{eqnarray} 
where we have used the notations $A$, $S$, and $L$ for three types of
mixing matrices, respectively. The second of three types are described
by the matrices with one of mixing angles being zero and the others
being maximal:
\begin{eqnarray}
  B &=& \left(
    \matrix{ \frac{1}{\sqrt{2}} & \frac{1}{\sqrt{2}} & 0 \cr
      -\frac{1}{2} & \frac{1}{2} & \frac{1}{\sqrt{2}} \cr
      \frac{1}{2} & -\frac{1}{2} & \frac{1}{\sqrt{2}} \cr} \right), 
  \quad\qquad
  \matrix{ s_{12}=1/\sqrt{2} \hfill\cr
    s_{13}=0 \hfill\cr
    s_{23}=1/\sqrt{2} \hfill\cr} \\[3mm]
  H &=& \left(
    \matrix{ \frac{1}{2} & \frac{1}{2} & \frac{1}{\sqrt{2}} \cr
      -\frac{1}{\sqrt{2}} & \frac{1}{\sqrt{2}} & 0 \cr
      -\frac{1}{2} & -\frac{1}{2} & \frac{1}{\sqrt{2}} \cr} \right), 
  \quad\qquad
  \matrix{ s_{12}=1/\sqrt{2} \hfill\cr
    s_{13}=1/\sqrt{2} \hfill\cr
    s_{23}=0 \hfill\cr} \\[3mm]
  N &=& \left(
    \matrix{ \frac{1}{\sqrt{2}} & 0 & \frac{1}{\sqrt{2}} \cr
      -\frac{1}{2} & \frac{1}{\sqrt{2}} & \frac{1}{2} \cr
      -\frac{1}{2} & -\frac{1}{\sqrt{2}} & \frac{1}{2} \cr} \right).
  \quad\qquad 
  \matrix{ s_{12}= 0 \hfill\cr
    s_{13}=1/\sqrt{2} \hfill\cr
    s_{23}=1/\sqrt{2} \hfill\cr}
\end{eqnarray} 
The threefold maximal mixing~\cite{threefold} and the unit matrix are
also added into our analysis:
\begin{eqnarray}
  T &=& \left(
    \matrix{ \frac{1}{\sqrt{3}} & \frac{1}{\sqrt{3}} &
      \frac{1}{\sqrt{3}} e^{-i\delta} \cr
      -\frac{1}{2}-\frac{1}{2\sqrt{3}}e^{i\delta} & \frac{1}{2}
      -\frac{1}{2\sqrt{3}}e^{i\delta} & \frac{1}{\sqrt{3}} \cr
      \frac{1}{2}-\frac{1}{2\sqrt{3}}e^{i\delta} & -\frac{1}{2}
      -\frac{1}{2\sqrt{3}}e^{i\delta} & \frac{1}{\sqrt{3}} \cr}
    \right),
  \quad\qquad 
  \matrix{ s_{12}=1/\sqrt{2} \hfill\cr
    s_{13}=1/\sqrt{3} \hfill\cr
    s_{23}=1/\sqrt{2} \hfill\cr}
\end{eqnarray}
\begin{eqnarray}
  I &=& \left( 
    \matrix{ 1 & 0 & 0 \cr
      0 & 1 & 0 \cr
      0 & 0 & 1 \cr} \right).
  \quad\qquad 
  \matrix{ s_{12}= 0 \hfill\cr
    s_{13}=0 \hfill\cr
    s_{23}=0 \hfill\cr}
\end{eqnarray}
In addition to these, the so-called democratic
mixing~\cite{Democratic} is examined since this mixing pattern is
rather different from the above ones and might be derived from
well-motivated underlying theories:
\begin{eqnarray}
  D &=& \left(
    \matrix{ \frac{1}{\sqrt{2}} & \frac{1}{\sqrt{6}} &
      \frac{1}{\sqrt{3}} \cr
      -\frac{1}{\sqrt{2}} & \frac{1}{\sqrt{6}} & \frac{1}{\sqrt{3}}\cr 
      0 & -\frac{2}{\sqrt{6}} & \frac{1}{\sqrt{3}} \cr} \right).
  \quad\qquad
  \matrix{ s_{12}=1/2 \hfill\cr
    s_{13}=1/\sqrt{3} \hfill\cr
    s_{23}=1/\sqrt{2} \hfill\cr}
\end{eqnarray}
Note that if one of the matrix elements of $U_E$ ($U_\nu$) is zero,
one can take $P'$ ($\overline P'$) as a unit matrix without loss of
generality. The phase $\delta_E$ ($\delta_\nu$) can be absorbed into a
redefinition of $Q$ and $P''$ ($\overline P''$). This fact is easily
understood in view of the Jarlskog parameter~\cite{J} which measures
the sizes of $CP$ violation: in case that one (or more) matrix element
is zero, the Jarlskog parameter is vanished. Accordingly, the phase
factors $\delta_E$ and $\delta_\nu$ are included only in the type $T$
matrix.

With the above mixing matrices at hand, we have 81 combinations of
matrices for $U_{\rm MNS}$, in which the phases $\alpha$, $\beta$,
$\delta_E$, and $\delta_\nu$ are taken to be free parameters. We
examine the MNS matrices referring to the phenomenological constraints
coming from the atmospheric neutrino experiments. The Chooz
experiment~\cite{CHOOZ} also provides a useful guide for the
classification of mixing matrices, in particular, for 
the $(U_{\rm MNS})_{e3}$ element. On the other hand, as we mentioned
in the introduction, the solar neutrino problem may be solved with
both large and small mixing angle solutions, and we will deal with it
as predictions of each case of 81 combinations for $U_{\rm MNS}$. In
what follows, we take a convention where the mixing between the labels
2 and 3 is relevant to the atmospheric neutrinos and the mixing
between the labels 1 and 2 to the solar neutrino problem. After all,
we find that the 81 mixing patterns are classified into the following
five categories:
\begin{itemize}
\setlength{\itemsep}{0mm}
\item Class 1: small mixing for atmospheric neutrinos 
\item Class 2: large value of $(U_{\rm MNS})_{e3}$ 
\item Class 3: small mixing for atmospheric neutrinos if 
  $(U_{\rm MNS})_{e3}\ll 1$ by phase tuning
\item Class 4: consistent with the experiments by phase tuning
\item Class 5: consistent with the experiments independently of phase
  values
\end{itemize}
Only classes 4 and 5 are consistent with the experimental data. Our
result of classification is summarized in Table~1. We have also
numerically checked the ``stability'' of our classification by
allowing the fluctuations of all mixing angles in the region of 
$\theta_{ij}=\theta_{ij}\pm 5^\circ$, both in the charged-lepton and
neutrino sectors. It is found that these fluctuations make no change
in the classification table.

In Table~1, we first notice that there are several exchanging
symmetries. In the neutrino side, the exchanges $A\leftrightarrow B$, 
$S\leftrightarrow I$, and $L\leftrightarrow H$ do not modify the
table. The existence of these symmetries is easily
understood. The above exchanges only reverse the predictions for
solutions to the solar neutrino problem (from large to small mixing
angle and vice versa), and so the classification table remains
unchanged. We also have a similar $T\leftrightarrow D$ symmetry. In
the charged-lepton side, the exchange $B\leftrightarrow N$ leaves the
classification unchanged. This is a bit curious symmetry. Because of
the constraint from the Chooz experiment, it is usually assumed that a
bi-maximal mixing matrix takes the form of type $B$. It is, however,
found here that the matrix $N$, which has a large 1-3 mixing, gives
exactly the same results as the matrix $B$ does. Unlike the neutrino
side, two types of matrices ($B$ and $N$) give the same predictions
even for the solar neutrino solutions (the 1-2 mixing angles). The
difference exists only in the values of phase factors which are
tuned. This fact would give a new possibility of model building for
the fermion masses and mixing.

Class 5 contains the following six mixing patterns: 
\begin{equation}
  (U_E,\,U_{\nu}) \;=\;
  (A,\,S),\;\; (A,\,I),\;\; (I,\,A),\;\; (I,\,B),\;\; (D,\,S),\;\;
  (D,\,I).
  \label{class5}
\end{equation}
There are essentially only three types of combinations due to the
exchanging symmetries stated above. As we will discuss in the next
section, it is interesting that these six combinations are certainly
predicted by the various models which have been proposed to account
for fermion mass hierarchies. Notice, however, that at this stage we
do not refer any particular structures of mass matrices but only
discuss the combinations of two unitary matrices $U_E$ and $U_\nu$,
combined with the phenomenological constraints on the rotation 
angles. The coincidence of these two approaches might show a profound
connection between the mass eigenvalues and mixing angles. Another
interesting point we find in Eq.~(\ref{class5}) is that the
``naturalness'', i.e., the absence of parameter tuning indicates that
the large 1-2 mixing relevant to the solar neutrino problem must come
from the neutrino side (except for the cases of democratic
mixing). This is naturally understood in view of the charged-lepton
masses, and indeed commonly seen in the literature. That is, in the
charged-lepton sector, the mass hierarchy between the first and second
generations is too large for the large-angle solar
solutions.\footnote{Note that the democratic mass matrix cannot
  explain the mass hierarchy between these lighter families. Moreover,
  small perturbations do not necessarily result in large 1-2 mixing.}
It should be noticed here that the same result is obtained only from a
viewpoint of mixing matrices. This may be again regarded as a sign of
deep connections between masses and mixing angles.

In the category of class 4, there are 31 patterns of mixing
matrices. These patterns require suitable choices of phase values to
be consistent with the experimental data. The result is summarized in
Table~2, where we present the values of mixing angles for atmospheric
neutrinos ($\sin^2 2\theta_{\rm atm}$) and for solar neutrinos 
($\sin^2 2\theta_\odot$) in case where $(U_{\rm MNS})_{e3}$ is set to
be minimum. For each combination, we also show the relevant phases
which are tuned to obtain the minimum value of 
$(U_{\rm MNS})_{e3}$. In some cases, the mixing angles 
$\sin^2 2\theta_{\rm atm}$ and $\sin^2 2\theta_\odot$ have 
uncertainties since there still exists phase degrees of freedom 
with the minimized values of $(U_{\rm MNS})_{e3}$. The mixing patterns
in class 4 need various numbers of phase tuning in order to obtain
experimentally suitable MNS matrices. For example, the types 
$(U_E,\,U_\nu)=(A,\,A)$ and $(A,\,B)$, which are often seen in the
literature, requires only one phase tuning to fix all the mixing
angles in $U_{\rm MNS}$ (see also the next section). Clearly, fewer
numbers of parameter tuning are preferable for higher
predictability. We find from Table~2 that the 8 combinations:
\begin{equation}
  (U_E,\,U_\nu) \;=\; (S,\,N),\;\; (S,\,D),\;\; (L,\,N),\;\; 
  (L,\,D),\;\; (B,\,L),\;\; (H,\,A),\;\; (H,\,D),\;\; (N,\,L), 
  \label{new}
\end{equation}
have the same predictability as $(A,\,A)$ and $(A,\,B)$; all the
mixing angles can be settled by only one phase tuning. Remarkably,
these combinations have not been discussed so far in the literature
and would provide new possibilities for constructing models where
fermion masses and mixing angles are properly reproduced.

%%%%%%%%%%%%%%%%%%%%%%%%%%%%%%%%%%%%%%%%%%%%%%%%%%%%%%%%%%%%%%%%%%%%%
%%%%%%%%%%%%%%%%%%%%%%%%%%%%%%%%%%%%%%%%%%%%%%%%%%%%%%%%%%%%%%%%%%%%%
\section{Texture of Lepton Mass Matrices}

The results in the previous section have been obtained independently
of any structures of lepton mass matrices and hence of the mass
spectrum of neutrinos. In this section, we discuss some implications
of the above results for the forms of lepton mass matrices. Note that
the mass textures we will discuss below are only examples among
various models which can lead to the same mixing patterns. There are
indeed infinite possibilities for mass matrices due to the remaining
freedom of mixing matrices, mass eigenvalues and their signs, the
particle property of neutrinos, etc. In the following discussion we do
not want to exhaust possible mass textures but, based on the previous
results, to show several examples which may correctly reproduce the
experimental data.

First we assume that neutrinos are Majorana particles, for
simplicity. The neutrino Majorana mass matrix $M_\nu$ is constructed 
by $U_\nu^* M_\nu^{\rm diag} U_\nu^\dagger$, where $M_\nu^{\rm diag}$
is the diagonal neutrino mass matrix but is not fully determined
even if the experimental data is given. In most cases below, we adopt
a hierarchy in the neutrino masses $M_\nu^{\rm diag}$. Moreover 
if one assumes the seesaw mechanism for tiny mass scales, $M_\nu$ 
is a low-energy effective mass matrix and a full mass matrix form may
be highly complicated. On the other hand, the charged-lepton mass
matrix is constructed by $U_E M_E^{\rm diag} R_E^\dagger$ where 
$M_E^{\rm diag}$ is the diagonal charged-lepton mass matrix and $R_E$
is the mixing matrix which rotates the right-handed charged-lepton
fields. Since $R_E$ is experimentally unknown, the charged-lepton mass
matrix is not uniquely reconstructed. In the following examples, we
assume $R_E=I$ or $U_E$, and take hierarchical mass eigenvalues which
may be parametrized by the Cabibbo angle $\lambda$.

Let us begin by discussing the mixing patterns in class 5. As noted
in the previous section, these mixing patterns have often appeared
in the literature. In other words, there are various models of lepton
mass matrices which lead to these mixing patterns. We overview the six
patterns in class 5. The first is the case $(U_E,\,U_\nu)=(A,\,S)$,
which predicts bi-maximal mixing for the MNS matrix. Assuming $R_E=I$,
we obtain a charged-lepton mass matrix $M_E$ and a Majorana mass
matrix $M_\nu$ as
\begin{eqnarray}
  M_E \,\propto\, \left( 
    \matrix{ & & \cr
     & \lambda^2 & 1 \cr
     & \lambda^2 & 1 \cr} \right), \qquad
  M_\nu \,\propto\, \left(
    \matrix{ \epsilon & \epsilon & \cr
     \epsilon & \epsilon & \cr
     & & 1 \cr} \right),
 \label{AS}
\end{eqnarray}
where $\epsilon$ is a small parameter and the blanks in the matrices
mean smaller entries. Such a type of texture has been derived, for 
example, in $SO(10)$ grand unified models~\cite{P1}. These models 
adopt the seesaw mechanism, and the source of large mixing in $M_\nu$
comes from the Dirac-type mass matrix of neutrinos, which is connected
with that of down quarks under the GUT symmetry.

The next pattern is the case $(U_E,\,U_\nu)=(A,\,I)$ that predicts
single maximal mixing for the MNS matrix, and it can be derived from,
for example,
\begin{eqnarray}
  M_E \,\propto\, \left( 
    \matrix{ & & \cr
     & \lambda^2 & 1 \cr
     & \lambda^2 & 1 \cr} \right), \qquad
  M_\nu \,\propto\, \left( 
    \matrix{ m_1 & & \cr
      & m_2 & \cr
      & & m_3 \cr} \right),
  \label{AI}
\end{eqnarray} 
where we have taken $R_E=I$. These mass matrices are indeed obtained
in $E_7$, $E_6$, and $SO(10)$ GUT models~\cite{P2}. In the GUT mass
textures (\ref{AS}) and (\ref{AI}), the large mixing in $M_E$ is
achieved by the mixing among the standard-model fields and extra
particles. The third example is $(U_E,\,U_\nu)=(I,\,A)$, which gives
single maximal mixing for the MNS matrix, and it leads to
\begin{eqnarray}
  M_E \,\propto\, \left( 
    \matrix{ \lambda^{4-5} & & \cr
     & \lambda^2 & \cr
     & & 1 \cr} \right), \qquad
  M_\nu \,\propto\, \left( 
    \matrix{ \epsilon & & \cr
      & 1 & 1 \cr
      & 1 & 1 \cr} \right),
\end{eqnarray} 
where $R_E=I$ is assumed.
This texture has been discussed, for example, in the $R$-parity
violating models~\cite{P3}. The fourth one is $(U_E,\,U_\nu)=(I,\,B)$,
which gives bi-maximal mixing. Assuming $R_E=I$, it gives
\begin{eqnarray}
  M_E \,\propto\, \left( 
    \matrix{ \lambda^{4-5} & & \cr
      & \lambda^2 & \cr
      & & 1 \cr} \right), \qquad
  M_\nu \,\propto\, \left( 
    \matrix{ & 1 & 1 \cr
      1 & & \cr
      1 & & \cr} \right).
\end{eqnarray} 
It has been shown that this texture follows from the radiative
generation mechanisms for neutrino masses~\cite{P4}.

The fifth and sixth patterns are a bit special since they depend on
the democratic lepton mass matrix~\cite{Democratic}, which usually
predicts $R_E=U_E$. The combination $(U_E,\,U_\nu)=(D,\,S)$ predicts
single maximal mixing for the MNS matrix, and gives
\begin{eqnarray}
  M_E \,\propto\, \left ( 
   \matrix{ 1 & 1 & 1 \cr
     1 & 1 & 1 \cr
     1 & 1 & 1 \cr} \right), \qquad
  M_\nu \,\propto\, \left( 
   \matrix{ 1 & \epsilon' & \cr
     \epsilon' & 1 & \cr
     & & 1+\epsilon \cr} \right).
 \label{DS}
\end{eqnarray}
On the other hand, $(U_E,\,U_\nu)=(D,\,I)$ has nearly bi-maximal
mixing, and gives
\begin{eqnarray}
  M_E \,\propto\, \left( 
    \matrix{ 1 & 1 & 1 \cr
      1 & 1 & 1 \cr
      1 & 1 & 1 \cr} \right), \qquad
  M_\nu \,\propto\, \left( 
    \matrix{ m_1 & & \cr
      & m_2 & \cr
      & & m_3 \cr} \right).
  \label{DI}
\end{eqnarray} 

We here again stress that all the above mixing patterns in class 5 are
allowed by the experimental data without any tuning of (sometimes
unphysical) phases $\alpha$, $\beta$, and $\delta_{E,\,\nu}$.

Next let us discuss the mixing patterns in class 4, where the presence 
of phase factors is essential for the MNS matrix to have the right
values of mixing angles. The 31 mixing patterns are classified into
this category, but only a few mass matrix models with these patterns
have been constructed. These patterns thus could provide potentially
useful textures of lepton mass matrices.

At first, we discuss the well-known example $(U_E,\,U_\nu)=(A,\,A)$
which leads to the following mass matrices by taking $R_E=I$:
\begin{eqnarray}
  M_E \,\propto\, \left( 
    \matrix{ & & \cr
      & \lambda^2 & 1 \cr
      & \lambda^2 & 1 \cr} \right), \qquad  
  M_\nu \,\propto\, \left(
    \matrix{ & & \cr
     & 1 & 1 \cr
     & 1 & 1 \cr} \right).
 \label{M41}
\end{eqnarray} 
This form often appears in the models with $U(1)$ flavor
symmetries~\cite{U1,vissani}. With this texture, the mixing angles at
leading order become
\begin{eqnarray}
  \theta_{\rm atm} \,=\, \frac{\beta-\alpha}{2},\qquad
  \theta_\odot \,=\, (U_{\rm MNS})_{e3} \,=\, 0,
  \label{E41}
\end{eqnarray}
where $\alpha$ and $\beta$ are the phase parameters in the matrix $Q$
(see Eq.~(\ref{Q})). This gives the SMA solution for the solar 
neutrino problem, and the constraint $(U_{\rm MNS})_{e3}\ll 1$ is also
satisfied. For the atmospheric neutrinos, however, one must tune the
phase values so that $\beta-\alpha \simeq \pi/2$. That is, due to the
presence of the phase matrix $Q$, the cancellation of two large mixing
angles from $U_E$ and $U_\nu$ can be avoided. Even though the
Super-Kamiokande result still allows about $10^\circ$ deviations from
the maximal mixing angle, the result can be explained only in about
20\% region of the whole phase parameter space of 
($\alpha$, $\beta$). This fact means that some amount of tuning of
phase parameters is indeed required to have right predictions. It
should be noted we numerically checked that this situation is
unchanged even if one includes $\pm 20^\circ$ fluctuations of the
mixing angles in $U_E$ and $U_\nu$.

Another pattern, for which the concrete models have been constructed,
is the case of $(U_E,\,U_\nu)=(A,\,B)$. Assuming $R_E=I$, it leads to
the mass matrix form
\begin{eqnarray}
  M_E \,\propto\, \left( 
    \matrix{ & & \cr
      & \lambda^2 & 1 \cr
      & \lambda^2 & 1 \cr} \right), \qquad  
  M_\nu \,\propto\, \left( 
    \matrix{ & 1 & 1 \cr
      1 & &  \cr
      1 &  & \cr} \right).
  \label{P42}
\end{eqnarray}
These textures have been discussed in Ref.~\cite{Shafi}. It is also
pointed out in Ref.~\cite{vissani} that this mixing pattern can be
predicted by the texture in Eq.~(\ref{M41}). The mixing angle 
$\theta_{\rm atm}$ is the same as in Eq.~(\ref{E41}), and a phase
combination $\beta-\alpha$ must be tuned so that one gets the
maximal mixing of atmospheric neutrinos.

As we stated in section 2, there are several new mixing patterns in 
class 4 which have not yet been discussed. Like the cases $(A,\,A)$
and $(A,\,B)$, the eight new patterns in Eq.~(\ref{new}) only need a
single phase tuning for fitting all the experimental data; the solar,
atmospheric, and long baseline neutrino experiments. Let us show an
example for the case $(U_E,\,U_\nu)=(S,\,N)$. This mixing pattern,
with $R_E=I$, gives the following form of mass matrices:
\begin{eqnarray}
  M_E \,\propto\, \left( 
    \matrix{  \lambda^{4-5} & \lambda^2 &  \cr
       \lambda^{4-5} & \lambda^2 & \cr   &  & 1 \cr} \right), \qquad  
  M_\nu \,\propto\, \left( 
    \matrix{ 2 & \sqrt{2} & \sqrt{2} \cr
      \sqrt{2} & 1+\epsilon & 1-\epsilon \cr
      \sqrt{2} & 1-\epsilon & 1+\epsilon \cr} \right).
  \label{P43}
\end{eqnarray} 
In this case, we have 
\begin{eqnarray}
  && \sin^2 2\theta_{\rm atm} \,=\, \sin^2 2\theta_\odot 
  \,=\, \left|\frac{1}{2}+\frac{1}{2\sqrt{2}}e^{i\alpha}\right|^2, 
  \nonumber \\
  && (U_{\rm MNS})_{e3} \,=\, 
   \left|\frac{1}{2}-\frac{1}{2\sqrt{2}}e^{i\alpha}\right|.
  \label{P43m}
\end{eqnarray}
Here we would like to emphasis that a single phase tuning of $\alpha$
ensures all the mixing angles to be consistent with the
experiments. A smaller value of $(U_{\rm MNS})_{e3}$ tuned by a phase
rotation automatically leads to larger mixing angles for solar and
atmospheric neutrinos. In this example, the mixing angle 
$\sin^2 2\theta_{\rm atm}$ might be a bit smaller than the
experimental bound from the atmospheric neutrino anomaly. One can,
however, easily get a proper MNS matrix if a few deviations from the
rigid values of mixing angles in $U_{E,\,\nu}$ are taken into
account. (Such deviations just correspond to those in the mass
matrices (\ref{P43}).) As we mentioned earlier, even with these
deviations, the classification is not changed and it is enough to tune
only one phase parameter. Notice that since the $(U_{\rm MNS})_{e3}$
mixing in Eq.~(\ref{P43m}) is close to the Chooz bound, this pattern
will be tested in the near future.

For the examples that need more than one phase tuning, we refer to the
models in~\cite{BB} which introduce the following types of mass
matrices:
\begin{eqnarray}
  M_E \,\propto\, \left( 
    \matrix{\lambda^{4-5} & \lambda^2 & 1 \cr
    \lambda^{4-5}  & \lambda^2 & 1 \cr
    \lambda^{4-5}  & \lambda^2 & 1 \cr} \right), \qquad  
  M_\nu \,\propto\, \left(
    \matrix{1 & 1 & 1 \cr
    1 & 1 & 1 \cr
    1 & 1 & 1 \cr} \right). 
\end{eqnarray} 
This corresponds to the mixing pattern $(U_E,\,U_\nu)=(T,\,T)$ or to
the special case $(U_E,\,U_\nu)=(D,\,D)$, where suitable MNS matrices
can also be obtained by phase tuning.

Including the above examples, we find that class 4 contains several
possible mixing patterns which no one has discussed so far (see
Table~2 and Eq.~(\ref{new})). Model construction utilizing such types
of textures may be worth performing.

Before closing this section, we note the connections of the 
low-energy Majorana neutrino mass matrices discussed above with those
at high-energy scale~\cite{ellis,HO}.\footnote{The
  renormalization-group equation of see-saw induced Majorana masses
  was first studied in Ref.~\cite{RGE}.} 
To discuss the stability of lepton flavor mixing against quantum
corrections, we need to determine the pattern of neutrino masses and
Majorana phases~\cite{HO,HOM}. For example, the neutrinos which are
degenerate in mass with the same phase sign may receive a considerable
change of flavor mixing structure~\cite{HO,HOS}. The mass matrices in
Eqs.~(\ref{AI}), (\ref{DS}), and (\ref{DI}), therefore, have a
possibility of changing the values of mixing angles during the
renormalization-group evolution. In particular, the mixing angles of
the democratic-type mass matrix (Eqs.~(\ref{DS}) and (\ref{DI})),
which is expected in the models with ${S_3}_L \times {S_3}_R$ or
$O(3)_L \times O(3)_R$ symmetries~\cite{Democratic}, might receive
large quantum modifications~\cite{HOMS}.

%%%%%%%%%%%%%%%%%%%%%%%%%%%%%%%%%%%%%%%%%%%%%%%%%%%%%%%%%%%%%%%%%%%%%
%%%%%%%%%%%%%%%%%%%%%%%%%%%%%%%%%%%%%%%%%%%%%%%%%%%%%%%%%%%%%%%%%%%%%
\section{Summary and Discussion}

To study the origins of the nearly maximal mixing of lepton flavor is
one of the most important issues in particle physics. We have examined
what types of mixing matrices of charged leptons and neutrinos can be
consistent with the neutrino experimental results. Our analyses in
section 2 do not depend on any details of underlying models, in
particular, of the mass matrix forms of charged leptons and
neutrinos. The results are hence independent of the mass spectrum and
property of neutrinos, for example, whether they are Dirac or Majorana
particles. As typical forms of charged-lepton and neutrino mixing
matrices, we have adopted nine types of unitary matrices, which
contain sources of large mixing angles and could be induced from some
underlying theories. We have then examined $9\times 9$ combinations of
mixing matrices and checked whether the resultant MNS mixing matrices
satisfy the phenomenological constraints from atmospheric and long
baseline neutrino experiments. In our analyses, the phase factors,
which cannot be absorbed into redefinitions of lepton fields, play
important roles.

As a result, we have found that there are various mixing patterns of
charged leptons and neutrinos for the MNS matrix with bi-maximal or
single maximal mixing. Among them, only six patterns are
experimentally allowed without any tuning of phase
values. Interestingly, these patterns are indeed derived from the 
concrete models which have been proposed to account for the fermion
mass hierarchy problem. The other patterns can give solutions to the
observed neutrino anomalies depending on the choices of phase
values. In this class of patterns, physically more significant mixing
patterns may be the ones which need fewer numbers of phase tuning to
have definite predictions. We have found that 10 combinations satisfy
this criterion; only a single phase tuning is required. They have not
been studied enough in lepton mass matrix models and will give new
possibilities of model-construction. Note that the tuned phases are 
not completely unphysical unlike in the quark sector, but some of them
are connected to Majorana phases and $CP$ violation phenomena in the
lepton sector. Combined with these effects, the improved measurements
of mixing angles $\sin^2 2\theta_\odot$ and $(U_{\rm MNS})_{e3}$ will
be important to select possible flavor mixing structures of leptons.

%%%%%%%%%%%%%%%%%%%%%%%%%%%%%%%%%%%%%%%%%%%%%%%%%%%%%%%%%%%%%%%%%%%%%
\vskip 0.5 cm
\subsection*{Acknowledgments}

We would like to thank T.~Yanagida for useful discussions and 
comments. We also thank the organizers and participants of Summer 
Institute 2000 held at Yamanashi, Japan, where a portion of this work 
was carried out. This work is supported by the Grant-in-Aid for
Science Research, Ministry of Education, Science and Culture, Japan
(No.~10640274, No.~12047220, No.~12740146, No.~12014208, No.~12047221, 
and No.~12740157).

%%%%%%%%%%%%%%%%%%%%%%%%%%%%%%%%%%%%%%%%%%%%%%%%%%%%%%%%%%%%%%%%%%%%%
\newpage
\setlength{\baselineskip}{23pt}

%%%%%%%%%%%%%%%%%%%%%%%%%%%%%%%%%%%%%%%%%%%%%%%%%%%%%%%%%%%%%%%%%%%%% 
\newpage
\begin{table}
\begin{center}
\def\arraystretch{2.4}
\begin{tabular}{|c||r|r|r|r|r|r|r|r|r|} \hline
  $U_E \backslash U_{\nu}$ &\quad $A$ &\quad $S$ &\quad $L$ &\quad $B$
  &\quad $H$ &\quad $N$ &\quad $T$ &\quad $I$ &\quad $D$ \\
  \hline\hline
  $A$ & {\bf 4} & {\bf 5} & 2 & {\bf 4} & 2 & 2 & 2 & {\bf 5} & 2 \\
  \hline 
  $S$ & 2 & 1 & 2 & 2 & 2 & {\bf 4} & {\bf 4} & 1 & {\bf 4} \\ \hline
  $L$ & 2 & 1 & 1 & 2 & 1 & {\bf 4} & {\bf 4} & 1 & {\bf 4} \\ \hline
  $B$ & 3 & 2 & {\bf 4} & 3 & {\bf 4} & 3 & {\bf 4} & 2 & {\bf 4} \\
  \hline 
  $H$ & {\bf 4} & 2 & 3 & {\bf 4} & 3 & {\bf 4} & {\bf 4} & 2 & 
  {\bf 4} \\ \hline
  $N$ & 3 & 2 & {\bf 4} & 3 & {\bf 4} & 3 & {\bf 4} & 2 & {\bf 4} \\
  \hline
  $T$ & {\bf 4} & 2 & {\bf 4} & {\bf 4} & {\bf 4} & {\bf 4} & {\bf 4}
  & 2 & {\bf 4} \\ \hline
  $I$ & {\bf 5} & 1 & 1 & {\bf 5} & 1 & 2 & 2 & 1 & 2 \\ \hline
  $D$ & 2 & {\bf 5} & 2 & 2 & 2 & {\bf 4} & {\bf 4} & {\bf 5} & 
  {\bf 4} \\ \hline
\end{tabular}
\caption{The classification of mixing patterns. The numbers denote the 
  categories defined in section 2.}
\end{center}
\end{table}

\begin{table}
\begin{center}
\def\arraystretch{2}
\tabcolsep=8pt
\begin{tabular}{|r|r|r|r|r|} \hline
  {$U_E$|$U_\nu$} & $\sin^2 2\theta_{\rm atm}$ & 
  ~~~~$\sin^2 2\theta_\odot$~~  & $(U_{\rm MNS})_{e3}$ & 
  {\small (No.~of) phases} \\ \hline\hline
  {$A$|$A$} & $0-1$ & 0 & 0 & (0) \\
  {$A$|$B$} & $0-1$ & 1 & 0 & (0) \\
  {$S$|$N$} & 0.73 & 0.73 & 0.15 & $\alpha$ \ (1) \\
  {$S$|$T$} & 8/9 & $1/4-1$ & 0 & $\alpha+\delta_\nu$ \ (1) \\
  {$S$|$D$} & 8/9 & 0 & 0 & $\alpha$ \ (1) \\
  {$L$|$N$} & 0.73 & 0.73 & 0.15 & $\beta$ \ (1) \\
  {$L$|$T$} & 8/9 & $1/4-1$ & 0 & $\beta+\delta_\nu$ \ (1) \\
  {$L$|$D$} & 8/9 & 3/4 & 0 & $\beta$ \ (1) \\
  {$B$|$L$} & 0.73 & 0.73 & 0.15 & $\beta$ \ (1) \\
  {$B$|$H$} & 0.73 & $0.23-0.96$ & 0.15 & $\beta$ \ (1) \\
  {$B$|$T$} & 8/9 & $1/4-1$ & 0 & $\alpha$, \ $\beta$ \ (2) \\
  {$B$|$D$} & 8/9 & 15/16 & 0 & $\alpha$, \ $\beta$ \ (2) \\
  {$H$|$A$} & 0.73 & 0.73 & 0.15 & $\alpha-\beta$ \ (1) \\
  {$H$|$B$} & 0.73 & $0.23-0.96$ & 0.15 & $\alpha-\beta$ \ (1) \\
  {$H$|$N$} & 1 & 1 & 0 & $\alpha$, \ $\beta$ \ (2) \\
  {$H$|$T$} & 8/9 & $1/16-1$ & 0 & $\alpha$, \ $\beta$ \ (2) \\
  {$H$|$D$} & 8/9 & 15/16 & 0 & $\alpha-\beta$ \ (1) \\ \hline
\end{tabular}
\caption{The mixing patterns in class 4. The values of mixing angles 
are shown in the case that $(U_{\rm MNS})_{e3}$ is minimal (the minimum 
values are also shown in the table). The last column denotes the
(number of) relevant phases which are needed for 
tuning $(U_{\rm MNS})_{e3}$. The uncertainties in 
$\sin^2 2\theta_{\rm atm}$ and $\sin^2 2\theta_\odot$ are fixed by
additional phase tunings.}
\end{center}
\end{table}

\begin{table}
\begin{center}
\def\arraystretch{2}
\tabcolsep=8pt
\begin{tabular}{|r|r|r|r|r|} \hline
  {$U_E$|$U_\nu$} & $\sin^2 2\theta_{\rm atm}$ & 
  ~~~~$\sin^2 2\theta_\odot$~~  & $(U_{\rm MNS})_{e3}$ & 
  {\small (No.~of) phases} \\ \hline\hline
  {$N$|$L$} & 0.73 & 0.73 & 0.15 & $\beta$ \ (1) \\
  {$N$|$H$} & 0.73 & $0.23-0.96$ & 0.15 & $\beta$ \ (1) \\
  {$N$|$T$} & 8/9 & $1/4-1$ & 0 & $\alpha$, \ $\beta$ \ (2) \\
  {$N$|$D$} & 8/9 & 15/16 & 0 & $\alpha$, \ $\beta$ \ (2) \\
  {$T$|$A$} & 1 & 8/9 & 0 & $\delta_E$, \ $\alpha-\beta$ \ (2) \\
  {$T$|$L$} & 1 & 8/9 & 0 & $\delta_E$, \ $\beta$ \ (2) \\
  {$T$|$B$} & 1 & $1/9-1$ & 0 & $\delta_E$, \ $\alpha-\beta$ \ (2) \\
  {$T$|$H$} & 1 & $1/9-1$ & 0 & $\delta_E$, \ $\beta$ \ (2) \\
  {$T$|$N$} & $8/9-1$ & 8/9 & 0 & $\alpha$, \ $\beta$ \ (2) \\
  {$T$|$T$} & $0-1$ & $0-1$ & 0 & $\alpha+\delta_\nu$, \ 
  $\beta+\delta_\nu$ \ (2) \\
  {$T$|$D$} & $0-1$ & $0-1$ & 0 & $\alpha$, \ $\beta$ \ (2) \\
  {$D$|$N$} & $1/36-0.96$ & 0.73 & 0.15 & $\alpha$ \ (1) \\
  {$D$|$T$} & $0-1$ & $1/4-1$ & 0 & $\alpha+\delta_\nu$ \ (1) \\
  {$D$|$D$} & $0-1$ & 0 & 0 & $\alpha$ \ (1) \\ \hline
\end{tabular}
\vskip 3mm

Table 2 (continued.)
\end{center}
\end{table}

\end{document}